\documentclass[a4paper,12pt]{article}
\title{Index for Orbifold Quiver Gauge Theories}
\author{Yu Nakayama}
\usepackage{amsmath}
\usepackage{amssymb}
\usepackage{graphicx}
\def\drawbox#1#2{\hrule height#2pt
        \hbox{\vrule width#2pt height#1pt \kern#1pt
              \vrule width#2pt}
              \hrule height#2pt}

\def\Fund#1#2{\vcenter{\vbox{\drawbox{#1}{#2}}}}
\def\Asym#1#2{\vcenter{\vbox{\drawbox{#1}{#2}
              \kern-#2pt       
              \drawbox{#1}{#2}}}}

\def\funda{\Fund{6.5}{0.4}}

\def\symm{\funda\kern-0.4pt\funda}

\allowdisplaybreaks[0]

\newcommand{\cN}{{\cal N}}

\newcommand{\sectiono}[1]{\section{#1}\setcounter{equation}{0}}

\begin{document}

\begin{titlepage}
\thispagestyle{empty}
\begin{flushright}
UT-05-22\\
hep-th/0512280\\
\end{flushright}

\vskip 1.5 cm

\begin{center}
\noindent{\textbf{\LARGE{\\\vspace{0.5cm} Index for Orbifold Quiver Gauge Theories
}}} 
\vskip 1.5cm
\noindent{\large{Yu Nakayama}\footnote{E-mail: nakayama@hep-th.phys.s.u-tokyo.ac.jp}}\\ 
\vspace{1cm}
\noindent{\small{\textit{Department of Physics, Faculty of Science, University of 
Tokyo}} \\ \vspace{2mm}
\small{\textit{Hongo 7-3-1, Bunkyo-ku, Tokyo 113-0033, Japan}}}
\end{center}
\vspace{1cm}
\begin{abstract}
We compute the index for orbifold quiver gauge theories. We compare it with the results  obtained from the type IIB supergravity (superstring) on $AdS_5 \times S^5/\Gamma$.

\end{abstract}

\end{titlepage}


\sectiono{Introduction}\label{sec:1}
The concept of the index in four dimensional super conformal field theories (SCFTs) compactified on $S^3 \times R$ was introduced in \cite{Romelsberger:2005eg,Kinney:2005ej}. While the index is defined in a similar way to the usual Witten index \cite{Witten:1982df} on $R^4$ (or rather $T^3 \times R$), it contains much richer information about the SCFT because not only vacuum states but also all the short multiplets of the SCFT will contribute. Therefore the study on the index is important in order to understand the structure of four dimensional SCFTs  and their classifications. In particular, it will lead to a nontrivial check of the  AdS-CFT correspondence (see \cite{Aharony:1999ti} and references therein).

The (twisted) Witten index for $\cN=1$ SCFT on $S^3 \times R$ \cite{Romelsberger:2005eg,Kinney:2005ej} is defined as 
\begin{eqnarray}
\mathcal{I}^W(t,y) = \mathrm{Tr} (-1)^Fe^{-\beta \Delta} t^{2(E+j_1)}y^{2j_2} \ , \label{Witten}
\end{eqnarray}
where $E$ is the energy (or conformal dimension via conformal mapping from $S^3 \times R$ to $R^4$) and $j_1, j_2$ denote the spin quantum number corresponding to the rotation around $S_3$ whose isometry is $SU(2)_1\times SU(2)_2$. The regularizing factor $\Delta$ is given by the anti-commutator of a specific supercharge\footnote{We use the notation of $\dagger$ as the BPZ conjugation in the radial quantization scheme: if we go back to $R^4$ by conformal mapping, the BPZ conjugation of the SUSY transformation is given by the superconformal transformation: $Q_{\alpha}^\dagger = S_{\alpha}$.}
\begin{eqnarray}
\Delta = 2\{Q^{-\frac{1}{2} \dagger}, Q^{-\frac{1}{2}}\} = E-2j_1 - \frac{3}{2}r\ ,
\end{eqnarray}
which should be positive definite from the unitarity. Here we denote $r$ as the $U(1)$ R symmetry of the $\mathcal{N}=1 $ superconformal algebra $SU(2,2|1)$ \cite{Flato:1983te,Dobrev:1985qv}.

This index has several important properties:
\begin{itemize}
	\item Only contribution to the index comes from the states with $\Delta = 0$, or short multiplets of the superconformal algebra. As an immediate consequence, the index does not depend on $\beta$.
	\item It does not depend on the continuous parameter of the theory such as the coupling constants. It is also invariant under any (marginal) deformation of the SCFT unless it breaks the superconformal invariance.
	\item The index \eqref{Witten} is unique in the sense that all the information of possible indices protected from the superconformal algebra alone is encoded in \eqref{Witten} (see \cite{Kinney:2005ej}).
\end{itemize}

In this paper, we study the quiver gauge theories\footnote{See \cite{Halpern:1975yj} for an earlier study on the array of gauge theories.} that are obtained when we place $N$ D-branes at the origin of the orbifold $C^3/\Gamma$ \cite{Douglas:1997de}. If $\Gamma \subset SU(2)$, the resulting theory possesses $\cN=2$ supersymmetry and if $\Gamma \subset SU(3)$, it possesses $\cN=1$ supersymmetry. The orbifold quiver SCFTs have a fixed line of coupling constants expected from the Leigh-Strassler-like argument \cite{Leigh:1995ep}, (one of) which corresponds to the dilaton expectation value in the AdS dual. Through this fixed line, the SCFTs are continuously connected to free gauge theories by a marginal deformation. Since the index does not depend on such deformations, we can compute the index in the free gauge theory limit, which simplifies the practical computation.

On the other hand, the orbifold quiver gauge theory has an AdS-CFT dual description defined as the type IIB superstring theory on $AdS_5 \times S^5/\Gamma$ \cite{Kachru:1998ys,Lawrence:1998ja}. Therefore, if the AdS-CFT correspondence is correct,  we can compute the index also from the weakly coupled supergravity on  $AdS_5 \times S^5/\Gamma$ in the large 't Hooft coupling limit.\footnote{In the following, we always take the large $N$ limit.} In order to verify this, we compute the index independently from the supergravity side in this paper and compare the results with the gauge side. In some orbifold theories, we further need to introduce contributions from the twisted sector of the string theory. We will see that the origin of the apparent difference between the gauge theory computation and the supergravity computation (in the untwisted sector) indeed results from the twisted sector.

The organization of the paper is as follows. In section 2, we compute the index for orbifold quiver gauge theories in the free gauge coupling limit. In section 3, we compute the same index in the large coupling limit by using the AdS-CFT correspondence, and compare the two results. In section 4, we present our conclusion and some future outlook.

\sectiono{Index for quiver gauge theories}\label{sec:2}
In this paper, we study the index for $\cN = 1$ or $\cN= 2$ orbifold quiver gauge theories compactified on $S_3 \times R$. We restrict ourselves to a class of simple orbifold action $\Gamma = Z_k \subset SU(3)$ or $Z_k \subset SU(2)$ for $\cN = 1$ or $\cN= 2$ respectively. The action of $Z_k$ on $C^3 = (X_1,X_2,X_3)$ is generated by 
\begin{eqnarray}
\omega = \left( \begin{array}{ccc}
   \exp(2\pi i /k) & 0 &0  \\
  0  & \exp(-2\pi i/k) &0 \\
   0 & 0 &1
  \end{array}
 \right)
\end{eqnarray}
in $SU(2)$ corresponding to $\cN=2$ gauge theories, and 
\begin{eqnarray}
\omega = \left( \begin{array}{ccc}
   \exp(2\pi m_1 i/k) & 0 &0  \\
  0  & \exp(2\pi m_2 i/k) &0 \\
   0 & 0 &\exp(2\pi m_3 i/k)
  \end{array}
 \right)
\end{eqnarray}
with $m_1+m_2+m_3 = 0$ (mod $k$) in $SU(3)$ corresponding to $\cN=1$ gauge theories. We label the latter orbifold as $(m_1,m_2,m_3)$. If we put $N$ D-branes at the origin of the orbifold $C^3/\Gamma$, the effective gauge theory living on such branes are described by the quiver gauge theories.

The $\cN =1$ orbifold quiver gauge theory is the $U(N)^{k}$ gauge theory with hypermultiplets in the bifundamental representations transforming as
\begin{eqnarray}
\oplus_{i=1}^{k} (1,\cdots, N_i, \bar{N}_{i+1},1,\cdots,1) 
\end{eqnarray}
with the usual $\cN=2$ superpotential terms.
The $\cN =1$ orbifold quiver gauge theory is the $U(N)^k$ gauge theory with the following chiral multiplets 
\begin{align}
Q_i^1 &\sim \oplus_{i=1}^{k}(1,\cdots,1,N_{i},1,\cdots,1,\bar{N}_{i+m_1},1,\cdots,1) \cr
Q_i^2 &\sim
\oplus_{i=1}^{k}(1,\cdots,1,N_{i},1,\cdots,1,\bar{N}_{i+m_2},1,\cdots,1) \cr
Q_i^3 &\sim \oplus_{i=1}^{k}(1,\cdots,1,N_{i},1,\cdots,1,\bar{N}_{i+m_3},1,\cdots,1)
\end{align}
together with the superpotential
\begin{eqnarray}
W = h\sum_i Q_1^i Q_2^{i+m_1} Q_3^{i+m_1+m_2} - Q_2^{i} Q_1^{i+m_1} Q_3^{i+m_1+m_2} \ .
\end{eqnarray}
At the tree level, we take $h = g_{YM}$ descended from the parent $\cN=4$ SYM theory.

To compute the index 
\begin{eqnarray}
\mathcal{I}^W = \mathrm{Tr}(-1)^F t^{2(E + j_1)} y^{2j_2} \ ,
\end{eqnarray}
for orbifold quiver gauge theories, we can take the free coupling limit $g_{YM}, h \to 0$ because the index does not depend on them. Then, the computation becomes drastically simplified by using  the matrix model technique introduced by \cite{Sundborg:1999ue,Aharony:2003sx} in order to compute the partition function (including the index) of free gauge theories compactified on $S_3 \times R$.

\begin{table}[tb]
\begin{center}
\begin{tabular}{c|c|c|c}
 Letters        & $(-1)^F[E,j_1,j_2] $& $r$ & representation\\
 \hline
 ${Q}$&     $  [1,0,0] $         &$2/3$ & $\bar{N}\times {N}$ \\
  $\bar{\psi}_Q$&     $  -[\frac{3}{2},\frac{1}{2},0] $&$1/3$ & $N\times \bar{N}$ \\
  \hline
   ${\Phi}$&     $  [1,0,0] $         &$2/3$ & adj \\
  $\bar{{\psi}}_\Phi$&     $  -[\frac{3}{2},\frac{1}{2},0] $ &$1/3 $&adj \\
\hline
$F_{++}$&     $ [2,1,0] $ &$0 $&adj \\
${\lambda}_{\pm} $&     $ -[\frac{3}{2},0,\pm\frac{1}{2}] $ &$1 $&adj \\
\hline
$\partial_{++}{\lambda}_{-} + \partial_{+-}{\lambda}_{+}=0  $&     $ [\frac{5}{2},\frac{1}{2},0] $ &$1 $&adj \\
\hline
$\partial_{+\pm} $&     $ [1,\frac{1}{2},\pm\frac{1}{2}] $ &$0 $&1 
 \end{tabular}
\end{center}
\caption{List of the letters contributing to the index (hence $\Delta = 0$). Adjoint chiral multiplets $\Phi$ only appear in $\cN = 2$ orbifold as an $\cN=2$ vector multiplet.}
\label{tab:1}
\end{table}%

In the $\cN=1$ quiver gauge theories, the building block of the letter partition functions (with $\Delta = 0$) are summarized in table \ref{tab:1}. The vector multiplet transforms as an adjoint representation, and the chiral multiplet transforms as a bifundamental representation. Now by utilizing the results obtained in  \cite{Sundborg:1999ue,Aharony:2003sx} the index can be computed as the following matrix integral over the holonomy (unitary) matrix $U$
\begin{eqnarray}
\mathcal{I}^W = \int \prod_{i}^k[dU]_i \exp \left(\sum_{ij}\sum_{n=1}^\infty \frac{1}{n} f_{ij}(t^n,y^n) \chi_{ij}(U^n)\right) \ , \label{eff}
\end{eqnarray}
where $\chi_{ij}(U)$ is the character for each single letter index $f_{ij}(t,y)$. Explicitly, for the adjoint representation we have $\chi_{ii}(U) = \mathrm{Tr} U_i\mathrm{Tr} U_i^{\dagger}$ (with no summation over $i$), and for the bifundamental representation we have $\chi_{ij}(U) = \mathrm{Tr} U_i \mathrm{Tr}U_j^\dagger $.\footnote{We use the notation that $\mathrm{Tr}$ denotes the trace over the fundamental representation.}

For example, let us consider the $Z_3$ orbifold quiver gauge theory \cite{Kachru:1998ys}. In this model, the effective action for the matrix integral in \eqref{eff} is given by
\begin{align}
-S_{eff} &= \sum_{n=1}^{\infty} \frac{1}{n} \left[\frac{2t^{6n}- t^{3n}(y^{n}+y^{-n})}{(1-y^{n}t^{3n})(1-y^{-n}t^{3n})}(\mathrm{Tr} U^n_1\mathrm{Tr} U_1^{\dagger n} + \mathrm{Tr} U^n_2\mathrm{Tr} U_2^{\dagger n} + \mathrm{Tr} U^n_3\mathrm{Tr} U_3^{\dagger n}) \right. \cr &+ \frac{3t^{2n}}{(1-y^nt^{3n})(1-y^{-n}t^{3n})} (\mathrm{Tr} U^n_1\mathrm{Tr} U_2^{\dagger n} + \mathrm{Tr} U^n_2\mathrm{Tr} U_3^{\dagger n} + \mathrm{Tr} U^n_3\mathrm{Tr} U_1^{\dagger n})  \cr &-\left. \frac{3t^{4n}}{(1-y^nt^{3n})(1-y^{-n}t^{3n})} (\mathrm{Tr} U^n_1\mathrm{Tr} U_3^{\dagger n} + \mathrm{Tr} U^n_2\mathrm{Tr} U_1^{\dagger n} + \mathrm{Tr} U^n_3\mathrm{Tr} U_2^{\dagger n})\right] \ .
\end{align}

Integration over the unitary matrices $U_i$ can be easily performed in the large $N$ limit, where we introduce the eigenvalue distribution $\rho_i(\theta)$ with $\int_{-\pi}^\pi \rho_i(\theta) d\theta = 2\pi $. In the low temperature limit $(t\to 0)$, the saddle point of the large $N$ matrix integral is given by the constant distribution $\rho_i(\theta) = 1$. Then the integration over the unitary matrices is reduced to Gaussian integral over Fourier modes of $\rho(\theta) = \sum_n \rho_n e^{in\theta}$ around the trivial saddle point ($\rho_0=1$, $\rho_{n\neq0} = 0$). Performing the Gaussian integral over $\rho_n$, we obtain the final result for general $Z_k \subset SU(3)$ orbifold with a label $(m_1,m_2,m_3)$:
\begin{eqnarray}
\mathcal{I}^W_{gauge/\Gamma;U(N)} = \prod_{n=1}^\infty \frac{(1-y^{-n}t^{3n})^k(1- y^nt^{3n})^k} { (1-t^{2l_1n})^{k/l_1}(1-t^{2l_2n})^{k/l_2}(1-t^{2l_3n})^{k/l_3}} \ ,
\end{eqnarray}
where $l_i = k/\mathrm{gcd}(k,m_i)$. 
In this computation, we encounter the determinant of a circulant matrix such as 
\begin{eqnarray}
\left|
 \begin{array}{ccc}
   1-\frac{2t^6-t^3(y+y^{-1})}{(1-yt^3)(1-y^{-1}t^3)} & \frac{-3t^2}{(1-yt^3)(1-y^{-1}t^3)}&\frac{3t^4}{(1-yt^3)(1-y^{-1}t^3)} \\
 \frac{3t^4}{(1-yt^3)(1-y^{-1}t^3)} & 1-\frac{2t^6-t^3(y+y^{-1})}{(1-yt^3)(1-y^{-1}t^3)}  &\frac{-3t^2}{(1-yt^3)(1-y^{-1}t^3)} \\
   \frac{-3t^2}{(1-yt^3)(1-y^{-1}t^3)} & \frac{3t^4}{(1-yt^3)(1-y^{-1}t^3)}&1-\frac{2t^6-t^3(y+y^{-1})}{(1-yt^3)(1-y^{-1}t^3)} 
  \end{array}
  \right| = \frac { (1-t^{6})^3}{(1-y^{-1}t^3)^3(1-yt^{3})^3} \ 
  \end{eqnarray}
in the case of $Z_3 \subset SU(3)$. Such computation can be systematically performed by using the formula for the circulant determinant presented in Appendix \ref{sec:A}.

Similarly, for $Z_k \subset SU(2)$ orbifold case, we obtain the result as
\begin{eqnarray}
\mathcal{I}^W_{gauge/\Gamma;U(N)} = \prod_{n=1}^{\infty} \frac{(1-y^{-n}t^{3n})^k(1-y^nt^{3n})^k} { (1-t^{2nk})^2(1-t^{2n})^k} \ .
\end{eqnarray}
 
However, this is not the end of the story. So far, we have considered the quiver gauge theory as $U(N)^k$ gauge theory for computational simplicity in the matrix integral. In order to compare the index from the gravity computation via AdS-CFT correspondence, we have to remove IR free $U(1)^{k-1}$ vector multiplets from the theory.\footnote{Often these $U(1)$ parts are even anomalous, and cannot survive at the IR.} The free $U(1)$ vector multiplet contributes to the index as
\begin{align}
\mathcal{I}_{U(1)}^W &= \exp\left( \sum_{n=1}^{\infty}\frac{1}{n} \frac{2t^{6n}-t^{3n}(y^{n}+y^{-n})}{(1-y^nt^{3n})(1-y^{-n}t^{3n})} \right) \cr
&= \prod_{n=1}^\infty (1-y^{-n}t^{3n})(1-y^{n}t^{3n}) \ .
\end{align}
Since these $U(1)$ parts are decoupled, the subtraction of the index is easily done just by dividing the $U(N)$ index by contributions of the $U(1)^{k-1}$ vector multiplets.\footnote{The diagonal $U(1)$ (corresponding to the singleton representation) can also be easily subtracted in a similar manner, but we will not do this, following \cite{Kinney:2005ej}.} The final expression for the index of the quiver gauge theory for $Z_k \subset SU(3)$ orbifold is
\begin{eqnarray}
\mathcal{I}^W_{gauge/\Gamma} = \prod_{n=1}^\infty \frac{(1-y^{-n}t^{3n})(1-y^nt^{3n})} { (1-t^{2l_1n})^{k/l_1}(1-t^{2l_2n})^{k/l_2}(1-t^{2l_3n})^{k/l_3}} \ . \label{final}
\end{eqnarray}
The index for $Z_k \subset SU(2)$ orbifold is similarly given by 
\begin{eqnarray}
\mathcal{I}^W_{gauge/\Gamma} = \prod_{n=1}^{\infty}\frac{(1-y^{-n}t^{3n})(1-y^nt^{3n})} { (1-t^{2nk})^2(1-t^{2n})^k} \ .
\end{eqnarray}

\sectiono{Comparison with gravity}\label{sec:2}
As discussed in the introduction, since the index is independent of the coupling constant of the quiver gauge theories, we can freely take the large coupling limit. Then according to the AdS-CFT correspondence, we should be able to compute the same index for orbifold quiver gauge theories from the type IIB supergravity on $AdS_5 \times S^5/\Gamma$.

To do this, we first begin with the single particle index for type IIB supergravity on $AdS_5\times S^5$ given by the formula \cite{Gunaydin:1984fk,Kinney:2005ej}\footnote{Here we are following the notation of \cite{Kinney:2005ej}: $R_2$ and $R_3$ denote the second and third Dynkin labels of $SU(4)$ R symmetry respectively.}
\begin{align}
\mathcal{I}_{sp} &= \mathrm{Tr}_{sp} (-1)^F t^{2(E+j_1)}y^{2j_2} v^{R_2}w^{R_3} \cr  &= \frac{t^2/w}{1-t^2/w} + \frac{vt^2}{1-vt^2} + \frac{t^2w/v}{1-t^2w/v} - \frac{t^3/y}{1-t^3/y} -\frac{t^3 y}{1-t^3 y} \ .
\end{align}
Now the corresponding single particle partition function {\it in the untwisted sector} can be obtained by projecting this single particle index onto the $\Gamma$ invariant subspace \cite{Oz:1998hr}. Explicitly, in the $Z_k \subset SU(3)$ orbifold case with a label $(m_1,m_2,m_3)$, we obtain 
\begin{align}
\mathcal{I}_{sp/\Gamma}(t,y) &= \frac{1}{k} \sum_{i=1}^k \left(\frac{\omega_i^{m_1}t^2}{1-\omega_i^{m_1}t^2} + \frac{\omega_i^{m_2}t^2}{1-\omega_i^{m_2}t^2} + \frac{\omega_i^{m_3}t^2}{1-\omega_i^{m_3}t^2} - \frac{t^3/y}{1-t^3/y} -\frac{t^3 y}{1-t^3 y} \right) \cr
&= \frac{t^{2l_1}}{1-t^{2l_1}} + \frac{t^{2l_2}}{1-t^{2l_2}} + \frac{t^{2l_3}}{1-t^{2l_3}} - \frac{t^3/y}{1-t^3/y} -\frac{t^3 y}{1-t^3 y}
\end{align}
with $l_i = k/\mathrm{gcd}(k,m_i)$, and $\omega_i$ denotes $k$-th root of unity.

The total contribution to the index from the untwisted sector is given by the multiparticle contribution of this single particle index:
\begin{eqnarray}
\mathcal{I}^W_{grav/\Gamma} = \exp\left[\sum_{n=1}^{\infty} \frac{1}{n} \mathcal{I}_{sp/\Gamma}(t^n,y^n) \right] = \prod_{n=1}^\infty \frac{(1-y^{-n}t^{3n})(1-y^nt^{3n})} { (1-t^{2l_1n})(1-t^{2l_2n})(1-t^{2l_3n})} \ .
\end{eqnarray}
In particular, when $k$ does not share a common divisor larger than one with any of $m_i$, the index is given by
\begin{eqnarray}
\mathcal{I}^W_{grav/\Gamma} = \prod_{n=1}^\infty \frac{(1-y^{-n}t^{3n})(1-y^nt^{3n})} { (1-t^{2nk})^3} \ ,
\end{eqnarray}
which is in good agreement with the result from the gauge theory \eqref{final}.

In more general situations where $k$ has a common divisor larger than one (including the $\cN=2$ quiver case), we have to supplement the contribution from the twisted sector. For example, in the $Z_k \subset SU(2)$ orbifold case, the orbifold action has a fixed locus $S^1 \subset S^5$, and there exist twisted Kaluza-Klein (KK) fields localized in the $S^1$, which yield dual states corresponding to the twisted sector.

Let us look at the $Z_k \subset SU(2)$ orbifold a little bit more in detail. The gauge theory computation for $\cN=2, \hat{A}_{k-1}$ quiver theories gives the index
\begin{eqnarray}
\mathcal{I}^W_{gauge/\Gamma} = \prod_{n=1}^\infty \frac{(1-y^{-n}t^{3n})(1-y^nt^{3n})} { (1-t^{2nk})^2(1-t^{2n})^k} \ ,
\end{eqnarray}
while the supergravity computation in the untwisted sector gives
\begin{eqnarray}
\mathcal{I}^W_{grav/\Gamma} = \prod_{n=1}^\infty \frac{(1-y^{-n}t^{3n})(1-y^nt^{3n})} { (1-t^{2nk})^2(1-t^{2n})} \ .
\end{eqnarray}
The difference between the two
\begin{eqnarray}
\prod_{n=1}^\infty \frac{1}{(1-t^{2n})^{k-1}}
\end{eqnarray}
is understood as the contribution from $k-1$ massless twisted fields on $AdS_5 \times S^1$. Indeed, one can see that the twisted KK modes corresponding to a massless tensor multiplet yield the corresponding single particle contribution to the index.\footnote{See \cite{Gukov:1998kk} for an explicit correspondence of the chiral primary operators.}

In a similar way, we expect contributions from the twisted sector in $\cN=1$ quiver gauge theories although there is no fixed point of $Z_k \subset SU(3)$ action on $S^5$. Let $m_1$ has the greatest common divisor $q_1$ larger than one with $k$. Then $(X_1,X_2,X_3) = (1,0,0)$ will return to itself after acting $l_1 = k_1/q_1$ applications of $Z_k$ action. In this way the corresponding twisted sector has a fixed point and results in massless states. These $-3 +\sum_i q_i$ massless twisted modes should be included in the computation of the index as is the case with the $\cN=2$ orbifold.\footnote{In \cite{Razamat:2002tm}, the number of the massless twisted sector is enumerated in comparison with the number of the exactly marginal deformations.}

While we do not attempt a thorough discussion of the KK reduction of these twisted states, we can study the contribution of these states from the difference of the index discussed above. Comparing the gravity computation in the untwisted sector with the gauge theory computation, we conjecture that the twisted sector contributes to the single particle index as
\begin{eqnarray}
\mathcal{I}_{sp/\Gamma;twisted} = \sum_{i=1}^3 (q_i-1)\frac{t^{2l_i}}{1-t^{2l_i}} 
\end{eqnarray}
 or to the corresponding multiparticle index as
\begin{eqnarray}
\mathcal{I}^W_{\Gamma;twisted} = \prod_{n=1}^{\infty} \prod_{i=1}^3 \frac{1}{(1-t^{2l_i n})^{q_i-1}} \ .
\end{eqnarray}
Note that the single particle contribution can be interpreted as a contribution (with different weight) from $-3+\sum_i q_i$ independent particles as expected. It would be interesting if one could derive the spectrum of the twisted sector from a direct KK compactification.

\sectiono{Conclusion}
In this paper, we computed the index for orbifold quiver gauge theories. These theories are connected to the free gauge theories through the exactly marginal deformation (fixed line of SCFT), so the computation of the index can be done in the free gauge theory limit. We compare the results with the gravitational computation from the weakly coupled type IIB supergravity on $AdS_5\times S^5/\Gamma$. When the theory does not have a massless twisted sector, these two approaches completely agree with each other, but in the more general situation, we need to add contributions coming from the twisted sector. We have decomposed the contribution from the twisted sector into independent single particle contributions. In the $\cN=1$ examples, each contribution is not identical, and it would be interesting to see its origin by studying the detailed spectrum of the twisted sector and its KK compactification.

In more general quiver gauge theories, we can consider the type IIB superstring theories on $AdS_5 \times X$ where $X$ is a Sasaki-Einstein manifold. In addition to the orbifold of $S^5$ considered here, there are infinitely many examples of Sasaki-Einstein manifold ($T^{1,1}$ \cite{Klebanov:1998hh}, $Y^{p,q}$ \cite{Gauntlett:2004yd}, $L^{p,q,r}$ \cite{Cvetic:2005ft} for example), which give a more complicated AdS-CFT correspondence. A crucial difference in this case is that the dual SCFTs are not continuously connected to the free gauge theory through a marginal deformation, where the chiral multiplets obtain nontrivial anomalous dimensions. Thus, it is quite a nontrivial issue whether we can apply the matrix model technique mostly developed for free gauge theories.

Nevertheless, we can compute the index for these theories from the weakly coupled type IIB supergravity. Moreover, we expect there is no nontrivial contribution from the twisted sector (at least in the large $N$ limit) because the Sasaki-Einstein manifold under consideration is smooth. It would be interesting to study the index from the KK spectrum on $AdS_5\times X$ \cite{Ceresole:1999zs,Kihara:2005nt,Oota:2005mr}, and gain some insight into the structure of the corresponding quiver gauge theories.


\section*{Acknowledgements}
The author would like to thank S.~J.~Rey and C.~Romelsberger for fruitful discussions. He also acknowledges their hospitality in his stay at APCTP and Seoul national university, where a part of the work was done.
This research is supported in part by JSPS Research Fellowships
for Young Scientists.
       
\appendix\sectiono{Circulant Determinant}\label{sec:A}
The determinant of the circulant matrix is given by 
\begin{equation}
\left|
  \begin{array}{ccccc}
   x_{1} & x_2 &x_3 &\ldots& x_{n} \\
  x_{n}  & x_1 &x_2 &\ldots&  x_{n-1}  \\
   x_{n-1} & x_n &x_1& \ldots & x_{n-2} \\
   \vdots & \vdots& \vdots & \ddots &\vdots \\
 x_2 & x_3 & x_4 & \ldots & x_1 
  \end{array}
 \right| = \prod^n_{j=1} (x_1 + x_2 \omega_j + x_3\omega_j^3 + \cdots + x_n \omega_j^{n-1}) \ ,
\end{equation}
where $\omega_j$ is the $n$-th root of unity.

\bibliographystyle{utcaps}
\bibliography{orbifold}

\providecommand{\href}[2]{#2}\begingroup\raggedright\begin{thebibliography}{10}

\bibitem{Romelsberger:2005eg}
C.~Romelsberger, ``An index to count chiral primaries in N = 1 d = 4
  superconformal field theories,''
\href{http://www.arXiv.org/abs/hep-th/0510060}{{\tt hep-th/0510060}}.

\bibitem{Kinney:2005ej}
J.~Kinney, J.~Maldacena, S.~Minwalla, and S.~Raju, ``An index for 4 dimensional
  super conformal theories,''
\href{http://www.arXiv.org/abs/hep-th/0510251}{{\tt hep-th/0510251}}.

\bibitem{Witten:1982df}
E.~Witten, ``CONSTRAINTS ON SUPERSYMMETRY BREAKING,'' {\em Nucl. Phys.} {\bf
  B202} (1982)
253.

\bibitem{Aharony:1999ti}
O.~Aharony, S.~S. Gubser, J.~M. Maldacena, H.~Ooguri, and Y.~Oz, ``Large N
  field theories, string theory and gravity,'' {\em Phys. Rept.} {\bf 323}
  (2000) 183--386,
\href{http://www.arXiv.org/abs/hep-th/9905111}{{\tt hep-th/9905111}}.

\bibitem{Flato:1983te}
M.~Flato and C.~Fronsdal, ``REPRESENTATIONS OF CONFORMAL SUPERSYMMETRY,'' {\em
  Lett. Math. Phys.} {\bf 8} (1984)
159.

\bibitem{Dobrev:1985qv}
V.~K. Dobrev and V.~B. Petkova, ``ALL POSITIVE ENERGY UNITARY IRREDUCIBLE
  REPRESENTATIONS OF EXTENDED CONFORMAL SUPERSYMMETRY,'' {\em Phys. Lett.} {\bf
  B162} (1985)
127--132.

\bibitem{Halpern:1975yj}
M.~B. Halpern and W.~Siegel, ``ELECTROMAGNETISM AS A STRONG INTERACTION,'' {\em
  Phys. Rev.} {\bf D11} (1975)
2967.

\bibitem{Douglas:1997de}
M.~R. Douglas, B.~R. Greene, and D.~R. Morrison, ``Orbifold resolution by
  D-branes,'' {\em Nucl. Phys.} {\bf B506} (1997) 84--106,
\href{http://www.arXiv.org/abs/hep-th/9704151}{{\tt hep-th/9704151}}.

\bibitem{Leigh:1995ep}
R.~G. Leigh and M.~J. Strassler, ``Exactly marginal operators and duality in
  four-dimensional N=1 supersymmetric gauge theory,'' {\em Nucl. Phys.} {\bf
  B447} (1995) 95--136,
\href{http://www.arXiv.org/abs/hep-th/9503121}{{\tt hep-th/9503121}}.

\bibitem{Kachru:1998ys}
S.~Kachru and E.~Silverstein, ``4d conformal theories and strings on
  orbifolds,'' {\em Phys. Rev. Lett.} {\bf 80} (1998) 4855--4858,
\href{http://www.arXiv.org/abs/hep-th/9802183}{{\tt hep-th/9802183}}.

\bibitem{Lawrence:1998ja}
A.~E. Lawrence, N.~Nekrasov, and C.~Vafa, ``On conformal field theories in four
  dimensions,'' {\em Nucl. Phys.} {\bf B533} (1998) 199--209,
\href{http://www.arXiv.org/abs/hep-th/9803015}{{\tt hep-th/9803015}}.

\bibitem{Sundborg:1999ue}
B.~Sundborg, ``The Hagedorn transition, deconfinement and N = 4 SYM theory,''
  {\em Nucl. Phys.} {\bf B573} (2000) 349--363,
\href{http://www.arXiv.org/abs/hep-th/9908001}{{\tt hep-th/9908001}}.

\bibitem{Aharony:2003sx}
O.~Aharony, J.~Marsano, S.~Minwalla, K.~Papadodimas, and M.~Van~Raamsdonk,
  ``The Hagedorn / deconfinement phase transition in weakly coupled large N
  gauge theories,'' {\em Adv. Theor. Math. Phys.} {\bf 8} (2004) 603--696,
\href{http://www.arXiv.org/abs/hep-th/0310285}{{\tt hep-th/0310285}}.

\bibitem{Gunaydin:1984fk}
M.~Gunaydin and N.~Marcus, ``THE SPECTRUM OF THE S**5 COMPACTIFICATION OF THE
  CHIRAL N=2, D = 10 SUPERGRAVITY AND THE UNITARY SUPERMULTIPLETS OF U(2,
  2/4),'' {\em Class. Quant. Grav.} {\bf 2} (1985)
L11.

\bibitem{Oz:1998hr}
Y.~Oz and J.~Terning, ``Orbifolds of AdS(5) x S(5) and 4d conformal field
  theories,'' {\em Nucl. Phys.} {\bf B532} (1998) 163--180,
\href{http://www.arXiv.org/abs/hep-th/9803167}{{\tt hep-th/9803167}}.

\bibitem{Gukov:1998kk}
S.~Gukov, ``Comments on N = 2 AdS orbifolds,'' {\em Phys. Lett.} {\bf B439}
  (1998) 23--28,
\href{http://www.arXiv.org/abs/hep-th/9806180}{{\tt hep-th/9806180}}.

\bibitem{Razamat:2002tm}
S.~S. Razamat, ``Marginal deformations of N = 4 SYM and of its supersymmetric
  orbifold descendants,''
\href{http://www.arXiv.org/abs/hep-th/0204043}{{\tt hep-th/0204043}}.

\bibitem{Klebanov:1998hh}
I.~R. Klebanov and E.~Witten, ``Superconformal field theory on threebranes at a
  Calabi-Yau singularity,'' {\em Nucl. Phys.} {\bf B536} (1998) 199--218,
\href{http://www.arXiv.org/abs/hep-th/9807080}{{\tt hep-th/9807080}}.

\bibitem{Gauntlett:2004yd}
J.~P. Gauntlett, D.~Martelli, J.~Sparks, and D.~Waldram, ``Sasaki-Einstein
  metrics on S(2) x S(3),'' {\em Adv. Theor. Math. Phys.} {\bf 8} (2004)
  711--734,
\href{http://www.arXiv.org/abs/hep-th/0403002}{{\tt hep-th/0403002}}.

\bibitem{Cvetic:2005ft}
M.~Cvetic, H.~Lu, D.~N. Page, and C.~N. Pope, ``New Einstein-Sasaki spaces in
  five and higher dimensions,'' {\em Phys. Rev. Lett.} {\bf 95} (2005) 071101,
\href{http://www.arXiv.org/abs/hep-th/0504225}{{\tt hep-th/0504225}}.

\bibitem{Ceresole:1999zs}
A.~Ceresole, G.~Dall'Agata, R.~D'Auria, and S.~Ferrara, ``Spectrum of type IIB
  supergravity on AdS(5) x T(11): Predictions on N = 1 SCFT's,'' {\em Phys.
  Rev.} {\bf D61} (2000) 066001,
\href{http://www.arXiv.org/abs/hep-th/9905226}{{\tt hep-th/9905226}}.

\bibitem{Kihara:2005nt}
H.~Kihara, M.~Sakaguchi, and Y.~Yasui, ``Scalar Laplacian on Sasaki-Einstein
  manifolds Y(p,q),'' {\em Phys. Lett.} {\bf B621} (2005) 288--294,
\href{http://www.arXiv.org/abs/hep-th/0505259}{{\tt hep-th/0505259}}.

\bibitem{Oota:2005mr}
T.~Oota and Y.~Yasui, ``Toric Sasaki-Einstein manifolds and Heun equations,''
\href{http://www.arXiv.org/abs/hep-th/0512124}{{\tt hep-th/0512124}}.

\end{thebibliography}\endgroup

\end{document}